# Random lasers with tunable angular spectra for high contrast imaging

Huihui Shen[1], Yaoxing Bian[1], Junyan Li[1,2], Weinian Liu[1], Hanyi Xue[1] and Zhaona Wang[1,*]

[1]Department of Physics, Applied Optics Beijing Area Major Laboratory, Beijing Normal University, Beijing, China 100875.

[2]Department of Electrical & Computer Engineering, Texas A&M University, 188 Bizzell St, College Station, TX 77843

E-mail: zhnwang@bnu.edu.cn

**Abstract**: Random lasers with low spatial coherence have important potential applications in high quality imaging and displaying. Here, a random laser with tunable angular spectra is proposed and fabricated through directly coupling an asymmetric microcavity with a commercial optical fiber. The designed random laser demonstrates pumping direction-independence property in working threshold, and good storage stability. More importantly, the angular spectrum can be adjusted by pumping different microcavities on the fiber, the output random lasing with ring-shape in momentum space are used as illumination source for biological imaging. An excellent image with speckle-free and higher contrast is achieved due to the low spatial frequency-free of the illumination source even relative to that of the common white lighting. The results indicate that the designed random laser has great application values in the fields of high-resolution biological imaging, integration optics and smart sensing.

## Introduction

Laser has been widely used in imaging[1] and display[2] due to its high brightness, high coherence, high efficiency and good spectral control. However, high spatial coherence of laser will produce coherent speckle noise, which is not good for full-field imaging and display application[3,4]. Some passive techniques have been utilized to reduce the coherence of illumination lasing for suppressing the coherent noise[5,6]. As a result, the quality of lasing imaging and display has been greatly improved in speckle noise-suppress at the expense of time resolution by additional technique[7,8]. To better meet the requirement of lasing in real-time display technology, the active methods of eliminating the coherent noise are required for the next generation of lasing display[9,10]. Lasers with low spatial coherence have attracted lots of researchers' attention due to their unique transmission property and interesting physical phenomenon. Particularly, random laser with inherent randomness has demonstrated the excellent speckle noise suppression to be a very promising illumination source in lasing display and imaging[10-15]. However, as a key parameter of imaging quality, image contrast has been few reported for random lasing illumination. Thus, it is necessary to put more attention in image contrast. The corresponding methods are highly desired to improve the image contrast for random lasing illuminating system.

As a unconventional laser without optical cavity, random laser has unique properties of easy integration and low spatial coherence and great potential applications in imaging[16], photonic chip[17] and other fields[18]. The random scattering endows lasing with inherent randomness and further increases the difficulty of lasing field control. To better meet the application requirements of random laser, the random lasing field has been roughly controlled through emission wavelength[19], linewidth[20], polarization[10,21], and emission directions[10,22]. For example, by using the excellent guiding light property of optical fiber, fiber random laser was proposed through directly adding gain material in optical fibers to obtain a better emission directivity and higher output power [23,24]. Besides, the researchers have designed the random lasers on the end face[25] or side of the fiber[26,27] to control the emission direction of the random lasing. These works are mainly about the light field control of random lasing in real space. Recently, a ring-shaped random lasing in the momentum space has been designed through coupling random laser on the side of optical fiber, providing a new view of angular spectrum for random laser light field regulation[27]. As a critical parameter affecting the image quality[28], the angular spectra of the light source is often designed in certain distribution for microscopic imaging systems to further improve imaging contrast and resolution. In an ordinary optical microscopy, the high spatial frequency information of the imaged object is easily hidden by the low spatial frequency information due to their large intensity difference. It is a good way to suppress the low spatial frequency information by reducing and/or removing the in-plane low spatial frequency part in the angle-spectrum of the illumination light source such as the dark-field microscope[29] and the spatially incoherent ring illumination microscope[30]. When the object is illuminated by the structured light microscopy[31], the high spatial frequency information beyond the diffraction limit can enter the low spatial frequency area by frequency shift strategy and thus be collected by the objective lens to improve the imaging quality. It is necessary to design angular spectra-regulatable random laser for its further application in speckle-free, high-contrast, super-resolution microscopy imaging.

In this work, a random laser with tunable angular spectra (AS) distribution is proposed by directly coupling the asymmetric micro-cavities on the commercial optical fiber. The random lasing with weak components of in-plane low spatial frequency angular spectra (LSFAS) is achieved by coupling the processes of free transmission and coupling filter of the microcavity-fiber interface. And the ring-shape random lasing with LSFAS-free and high spatial frequency AS-tunability in momentum space is obtained through coupling filter and intensity modification induced by the refraction and reflection at the multi-interfaces along the optical fiber waveguide. The obtained random lasers have the low thresholds of 0.09 $MW/cm^2$ (yellow) and 0.30 $MW/cm^2$ (red), good emission directionality, pumping direction independence in threshold, emission polarization independence and long storage period. More importantly, the AS-tunable random lasing can be used as an excellent illumination source for suppressing coherent noise and improving the image contrast. The image contrast of a transparent biological sample is improved by 3.5 times by random lasing with weak LSFAS relative to that of white light imaging, respectively. The newly designed random laser with tunable AS distribution may extend the application of random laser in high-resolution biological imaging and real-time full-color display.

**Results and discussions**

The AS distribution-tunable random laser is designed with asymmetrical microcavity side-coupled on the fiber and shown in Fig. 1a. When a pulse lasing pumps the gain material, the photons are amplified through the multiple scattering of the particles in the microcavity. Random lasing will be radiated from the sample as the total gain is greater than the loss in the whole system as shown in Fig. 1a (real space). Part of the random lasing radiates directly into the air with free transmission. Part of radiation lasing is coupled into the optical fiber through the interface refraction and light guide of the optical fiber. For the free transmission (blue), the intensity of each AS component is theoretically uniformly distributed in photon momentum $\boldsymbol{k}$ space as shown in Fig.1b (dashed line in blue in $k_\parallel$-$k_z$ plane). For the lasing coupled into the optical fiber, the random lasing modes with small in-plane wavevector $\boldsymbol{k}_\parallel$ are forbidden and relative large in-plane wavevector $\boldsymbol{k}_\parallel$ modes can be guided along the optical fiber by using the microcavity-fiber interface refraction and optical fiber waveguide. The minimal radiation angles (in-plane wavevector $\boldsymbol{k}_\parallel$) of the random lasing from the fiber end are 39.91° (0.64$\boldsymbol{k}$) from the fiber core, 36.30° (0.59$\boldsymbol{k}$) from the cladding layer, and 32.58° (0.54$\boldsymbol{k}$) from the coating layer relative to the fiber axis, respectively. The detailed calculations are shown in Fig. S1a (Supporting Information). This AS-modulation process is called coupling filtering with an in-plane spatial frequency filtering function $F_{\mathrm{CF}}$:

$$F_{\mathrm{CF}}(k_\parallel, k_z) = \begin{cases} 1, & k_\parallel \geq k_{\min} \\ 0, & k_\parallel < k_{\min} \end{cases}. \tag{1}$$

Here, $k_{\min}$ is the corresponding in-plane minimum passing wavenumber. Considering the simple relation of in-plane wavenumber $k_\parallel$ and corresponding spatial frequency $f_\parallel = k_\parallel/2\pi$, we can conclude that this filter process can effectively filter out the lasing components of the in-plane LSFAS with $f_\parallel < f_{\min}$. Moreover, the lower cut-off spatial frequency $f_{\min}$ can be tuned by the refraction index of the corresponding guiding layer.

This property supplies an interesting way to flexibly modify the emission AS distribution in different spatial regions. As a result, AS distribution in $k_\parallel$-$k_z$ plane is modified by the proposed coupling filter process as shown in Fig. 1b (dashed line in black). Additionally, the amplitude transmission coefficients of different AS modes are different when these photons are travelling along the optical fiber due to the AS dependency of Fresnel coefficients and fiber length ($L$) dependency of total reflection times. According to the law of refraction and the Fresnel formula (Section **A**, Supporting Information), the AS distribution of the emission random lasing from the fiber end can be further regulated by designing different fiber lengths. This regulation process is named as intensity modulation with a function of $F_{\text{IM}}(k_\parallel, k_z, L)$, indicating that modifying the optical fiber length is a smart way for regulating the AS distribution. Based on the three processes mentioned above, the angle spectrum $A(k_\parallel, k_z)$ of random lasing emitting from the optical fiber end can be flexibly modulated by selectively using free transmission, coupling filtering and intensity modulation.

To better modify the emission AS, the asymmetric microcavity is asymmetrically fabricated on the side of an optical fiber near the left end (Fig. 1a). The AS component of the random lasing in the emission plane of end face can be obtained as

$$A(k_\parallel, k_z) = A_0(k_\parallel, k_z) F_{\text{CF}}(k_\parallel, k_z) F_{\text{IM}}(k_\parallel, k_z, L) + A_0(k_\parallel, k_z) F_{\text{FT}}(L). \tag{2}$$

Here, $F_{\text{FT}}(L) = 1/L^2$ is the transfer function for free transmission, $L$ represents the spacing between the random laser and emission plane. The AS distribution in the emission plane is determined and regulated by $L$. At the left emission plane named port 1 ($L$ is small), the random lasing is composed with those photons through free transmission and those modes undergoing coupling filter process through the optical fiber, i.e. $F_{\text{FT}} \to 1, F_{\text{IM}}(k_\parallel, k_z, L) \to 1$. The corresponding AS distribution includes the weak components of in-plane low spatial frequency (LSF) and relative strong components of high spatial frequency (HSF) as shown in $k_\parallel$-$k_z$ plane and $k_x$-$k_y$ plane in Fig.1b. At the right emission plane of port 2, the emitting random lasing undergoes a long transmission length of optical fiber ($L$=10 cm). The components with LSFAS are vanished through free transmission and filtered out through coupling filter, i.e. $F_{\text{FT}} \to 0$. And the corresponding LSFAS modes with $k_\parallel < k_{\min}$ are absent in the momentum space. Adding the intensity modulation, the AS distribution of the emission random lasing at port 2 is calculated as the right panel in Fig.1b, demonstrating a good LSFAS-free property. Considering almost rotational symmetry of the designed system, the AS distribution at port 2 demonstrates a ring-shape in momentum space in Fig. 1b ($k_x$-$k_y$ plane) as expected. The obvious difference of the AS distributions of random lasing from port 1 ($L\sim0$ cm, blue solid line) and from port 2 ($L=10$ cm, red solid line) indicates that the AS distribution of the proposed random laser can be flexibly regulated by controlling the optical fiber length.

In an optical microscope system, when an object is illuminated by parallel light with a large in-plane wave vector $k_\parallel$, part of the high-frequency information of the object will be shifted to the low-frequency area and further be detected by the system (Fig. S1b-c, Supporting Information) based on the Abbe imaging theory. The corresponding transmittance function of the imaging system is written as

$$H(k) = \int A(k_\parallel)H(k - k_\parallel)\mathrm{d}(k_\parallel) \quad . \tag{3}$$

Here $H$ is the frequency transmittance function of the objective lens and $A$ is the intensity of the different angular spectrum ($k_\parallel$). When the AS distribution of the illumination source is tuned with weak in-plane LSFAS components as port 1 mentioned above, the corresponding image contrast is thus improved due to the weak low spatial frequency information (Fig. S1d, Supporting Information). This is the fundamental of the high contrast imaging based on the AS-tunable random laser.

In the experiments, the AS-regulatable random laser is designed by coupling asymmetric microcavities on the optical fiber. A yellow random laser is first fabricated by selecting pyrromethene 567 (PM567) as the gain material in the polydimethylsiloxane (PDMS) and Ag nanoflowers (Ag NFs) with plenty of nanogaps as scattering media, demonstrating an optical fiber-integrated asymmetric microcavity in Fig. S2a (Supporting Information). The absorption and photoluminescence spectra of the obtained PM567 film demonstrates a strong absorption to the 532 nm pump pulses and an obvious fluorescence peak at about 558 nm in Fig. S2c (Supporting Information). The used Ag NFs in Fig. S2b (Supporting Information) have a diameter of about 0.5 μm and plenty of nanogaps and nanotips that can greatly enhance the local electromagnetic field and provide enough feedback for random systems[32,33]. The corresponding photoluminescence spectrum is effectively enhanced by the local surface plasmonic resonance of the Ag NFs[34] in Fig. S2c (Supporting Information). The fabricated random laser is pumped by a 532 nm pulsed laser as shown in Fig. S2d (Supporting Information). And the distance between the asymmetric microcavity and the fiber port can be controlled to achieve the control of angular spectra. The output random lasing can be divided into two channels for the output spectra analysis and AS distribution imaging or illuminating the object for imaging and display.

When the yellow random laser is pumped by 532 nm pulses, bright yellow random lasing is emitted from both ends of the fiber in Fig. 2a. The emission spot is circular in real space (Fig. 2b) and the ring-shape in momentum space in Fig. 2c as mentioned above. The unique angular spectrum distribution makes the obtained random lasing be an excellent illumination source in the microscopic imaging and display. The emission spectra of the random laser are shown in Fig. 2d at various pump power densities. When the pump power density is less than 0.20 MW/cm$^2$, only a wide spontaneous emission spectrum centered at 560 nm is observed. When the pump power density exceeds 0.20 MW/cm$^2$, several discrete narrow peaks with a linewidth of about 0.60 nm can be clearly observed, indicating that coherent resonance feedback is established in the sample[35]. The peak intensity and full width at half maximum (FWHM) evolution of the random lasing mode at 580 nm with the pump power density are shown in Fig. 2e. When the pump power density exceeds 0.20 MW/cm$^2$, the peak intensity increases rapidly, and the linewidth decreases sharply to sub-nanometer, indicating an operating threshold of around 0.20 MW/cm$^2$. The corresponding power spectrum Fourier transform (PFT) spectrum is shown in Fig. 2f based on the emission spectrum under a pump power density of 0.95 MW/cm$^2$. The effective optical cavity length $L_c$ is calculated as 108.7 μm by using the formula $p_m = mnL_c/\pi$ with $m$, $n, L_c$ and $p_m$ presenting the order of Fourier harmonics, the refractive index ($n$=1.40), the effective cavity length and the Fourier component[36], respectively.

The output directionality of the designed random laser is carefully characterized by demonstrating the emission intensity at various detection angles $\theta$ relative to the fiber axis when fixing a detection distance of 3.0 cm from the pump position (Fig. S3a, Supporting Information). At a pump power density of 0.95 MW/cm$^2$, the laser intensity along the fiber axis is much greater than that in other directions as the blue line in Fig. 2g and Fig. S3b (Supporting Information), indicating a good emission directivity of the designed random laser. Considering the symmetry in the central plane, the pump angle $β$ is defined as $β=π/2-α$ with an angle $α$ between the pump beam and the fiber axis (Fig. S3a, Supporting Information). The operation threshold of the random laser remains almost unchanged with changing the pump angle at a detection angle $θ=0$ in Fig. 2g (red line), demonstrating a pump angle-independency of the threshold. This is attributed to the sphere-like surface morphology of the microcavity supplying a perfect optical coupling channel for the pulses with different pump angles. The angle-independency is benefit to design a robust random laser in pump angle.

Polarization property of the emission random lasing is also analyzed by modifying the polarization of pump pulses through a half-wave plate under the condition of detection angle $θ=0$ and pump angle $β=0$ (Fig. S4a, Supporting Information). The output lasing integral intensity is the largest (smallest) when the polarization direction of pump beam is perpendicular (parallel) to the optical fiber axis in Fig. S4b (Supporting Information). When rotating the analyzer, the emission intensities of the random lasing from the fiber end facet are both almost unchanged under the pump polarization direction parallel (blue) and perpendicular (black) to the fiber axis (Fig. S4c-d, Supporting Information). Even when the 1/4 wave plate is introduced, these intensities are almost unchanged. This result indicates that the output random lasing is non-polarized lasing at $θ=0$. Interestingly, the coherent random lasing can still be measured and the threshold is almost unchanged when the sample is stored for one year (Fig. S5, Supporting Information). It shows that the proposed device has an excellent stability, providing a guarantee for the practical applications of random laser.

The effect of the gain dye concentration on the emission spectra of the random laser is also shown in Fig. S6a (Supporting Information). When the pump power density is 0.34 MW/cm$^2$, the center wavelength of the random lasing is red-shifted from 570 nm to 580 nm with increasing the dye concentration from 0.25 mg/mL to 3.0 mg/mL.

The results indicate that controlling the concentration of the dye is a good way to tuning the output wavelength of the designed random laser to better meet the actual application requirements. The output performances of the random lasers with different microcavity size are also carefully demonstrated in Fig. S6b and S6c (Supporting Information). As the microcavity diameter $d$ increases, the central wavelength of the random lasing shifts from 570 nm to 590 nm, demonstrating a clear red-shift phenomenon. This spectral redshift can be attributed that the larger the microcavity diameter means the larger gain region and stronger reabsorption[37,38]. At the same time, the emission intensity increases and operation threshold decreases with increasing $d$ (the minimum threshold is 0.09 MW/cm$^2$ in our experiments), demonstrating a good tunability of the emission performances.

The design approach of the AS-regulatable random laser can be easily extended to achieve various color random lasing. As an example, red random laser is realized as shown in Fig. 3. The gain material of the AS-regulated red random laser is 4-(Dicyanomethylene)-2-tert-butyl-6-(1,1,7,7-tetramethyljulolidin-4-yl-vinyl)-4H-pyran (DCJTB). Figure 3d is the emission spectra of the random lasing under the different pump power density of pump mode 1. At the same time, the FWHM sharply decreases to sub-nanometer with increasing the pump power density, indicating a coherent random lasing, demonstrating a low threshold of 0.30 MW/cm$^2$. The colorful random lasing assures its application in imaging and display as an illumination source.

Based on mechanism of the microcavity coupling the optical fiber, random lasers with tunable angular spectra are realized, by pumping the different microcavities integrating on the optical fiber as shown in Fig. 3a. Figure 3a is a geometric light path diagramare of the different AS distributions. The light field distribution on the z-axis behind the lens indicates that there is a long distance that intensity distribution is ring-shaped. Due to the large radius of the ring on the focal plane, the intensity distribution between the back focal plane and the imaging plane represents the intensity distribution of LSFAS modes. Figure 3b is the intensity pictures of the AS distribution in pump mode 1 and pump mode 2, corresponding to the marked position. It further shows that intensity of AS is different than under pump mode 1 or 2. In order to observe difference distribution of the AS between pump mode 1 and pump mode 2 intuitively, we take the gray value distribution of the longitudinal diameter of the picture at -4 cm as shown in Fig. 3c. It can be seen that the intensity of LSFAS modes in pump mode 2 is the lower. It shows that the AS of random lasing can be adjusted by simple mechanical movement, the pump source is closer to the fiber port of the microcavity, and the intensity of LSFAS modes is strength. Meanwhile, and they also have the good spectra and low threshold, as shown in Fig. 3d-e.

Furtherly, the prepared random lasers are used as the illumination source to demonstrates their potential in eliminating coherent speckle noise in microscopic imaging system for a biological sample including phloem, xylem, and pith. For comparison, 532 nm laser is first used as an illumination source with high spatial coherence. The corresponding images exhibit obvious speckle patterns (Figs. S7a-d, Supporting Information). These artificial intensity modulations make the cell boundary unclear even vanished. When the obtained random lasing is used as an illumination source, clear images of the whole biology object with random phase modulation are obtained with clear cell boundary and detail information in Figs. S7e-h (yellow, Supporting Information) and in Figs. S7i-k (red, Supporting Information). The results indicate that the designed random lasing can effectively eliminate the coherent artefacts and has a good application prospect in the field of full-field speckle-free biology imaging and display.

To order to reveal the application potential of the AS-tunable random laser in imaging contrast, the ability of improving the image quality of the fabricated random laser is approved by comparing the images which illuminated by the red random lasing with the different AS distributions in Fig. 4, respectively. Figure 4a are the intensity pictures of the AS distribution at -4 cm in pump mode 1 and pump mode 2, respectively. When these two sources with different AS distributions are used to illuminate a transparent sample, the random lasing source with less LSFAS has better imaging contrast in Fig. 5b. Figure 5c is the gray value distribution of the marked position in Fig. 5b. To quantitatively describe the imaging modulation depth, we define a parameter

$$K = \frac{G_b - G_{\min}}{G_b + G_{\min}},$$ $G_{\min}$ and $G_b$ is the local minimal value and the bedding value (the background field) of the depression in the grayscale image. the fluctuation range of the background field $G_b$ of the light source in the pump mode 1 is larger, which leads to a decrease in the contrast of the picture. We count the $G_b$ and variance of the background field at different structure positions in Fig. 5d. In pump mode 2, the average value (blue sphere ) of $G_b$ is around 0.851, and the variance (blue sphere) is around 0.00109. However, the average value of $G_b$ in pump mode 1 (white sphere ) is about 0.737 smaller than that in pump mode 2, and the variance (white sphere ) is about 0.00256 larger than that in pump mode 2. It's worth noting that the imaging contrast in pump mode 2 is higher than that in mode 1. To further characterize the change in contrast, corresponding to the background $G_b$, we take the corresponding $G_{\min}$ and get the corresponding statistical distribution of $K$. The modulation depth $K$ (blue sphere ) under pump mode 2 is always higher than that (white sphere ) under pump mode 1. Figure 5d shows that the corresponding improvement factor $\gamma = K_{RL}/K_{RL0}$ of the parenchyma cell in the bone marrow structure is always greater than 100%, which indicates that the designed random laser has less intensity of LSFAS and higher imaging contrast. These results further prove that the proposed AS-regulatable random laser have great advantages in the field of biological imaging and display with higher contrast.

By comparing the images of the three local parts which illuminated by the red random lasing under pump mode 2 and the broadband white source (Halogen Light Source, wave band:200~1700 nm) in Fig. 5a, the random lasing can achieve the same effect as white light illumination in suppressing speckle noise. More importantly, the images with the random lasing illumination have clearer cell boundary and a higher contrast than those illuminated by white source. The gray values of the five selected lines on the read picture with the pixel positions are shown in Fig. 5b. The obtained random lasing demonstrates an obvious superiority as an illumination source in imaging modulation depth. Particularly, when imaging for the high transparency region of the parenchyma cells, some detail information is vanishing under the broadband white light illumination. However, these structural details are obvious, when the sample is illuminated by random lasers with fewer LSFAS modes. The result indicates that the random laser has obvious advantages in transparent sample imaging. The *K* values are carefully calculated according to the gray maps at four different structures of the phloem, xylem, bubble and parenchyma cells in Fig. S8 (Supporting Information). The modulation depth *K* under the random lasing illumination (blue sphere) is always higher than that of the white light (white sphere). The corresponding improving factor $\gamma$ =$K_{RL}/K_0$ is always greater than 100% for four parts in Fig. 5c, demonstrating the better potential of the designed random lasing relative to the common white source. According to statistical analysis, the improving factor $\gamma$ of parenchyma cells in the bone marrow structure is 318%, the bubble is 135%, the phloem is 211%, and the xylem is 123%. These results further prove that the proposed random lasers have great advantages in the field of biological imaging and display with a higher contrast.

**Conclusions**

A random laser with tunable angular spectrum is proposed and realized by directly coupling asymmetric microcavity with the optical fiber, Which have great output performance and unique application in microscopic imaging system. Firstly, the obtained random lasers have the characteristics of easy integration, directional emission, pump angle-independency in threshold, unpolarization emission and good storage stability. This design method can be easily extended to achieve various color random lasing. The realized yellow and red random lasers with thresholds of 0.09 MW/cm$^2$, and 0.30MW/cm$^2$, respectively. Secondly, the random lasing also has low spatial coherence and the role of suppressing speckle noise in the microscope imaging system. More importantly, the position of the pump source can be controlled mechanically to control the distribution of the AS distribution of random lasing emitted from the end face of the fiber. The random lasing with less LSFAS modes can effectively reduce the background field in the image and thus improve the imaging contrast. These two characteristics of our designed random laser induce an excellent image of the biological sample. Finally, the image contrast under the obtained random lasing illuminations is improved by 318% for the parenchyma cells, 135% for the bubble, 211% for the phloem and 123% for xylem relative to that of the broadband white source illuminations. The results reveal that the proposed random lasers with tunable angular spectrum have great application values in the fields of speckle-free and high-contrast biological microimaging and display.

**Methods:**

**Random Laser fabrication.**

In the experiments, the used gain materials are 4-(Dicyanomethylene)-2-tert-butyl-6-(1,1,7,7-tetramethyljulolidin-4-yl-vinyl)-4H-pyran (DCJTB, Tokyi Chemical Industry) for red lasing and Pyrromethene 567 (PM567, Sigma) for yellow lasing. The home-made Ag nanoflowers (Ag NFs) are used as scattering particles for fabricating random lasers.

*Ag NF preparation*. $AgNO_3$ (99.9999%, Aldrich), polyvinylpyrrolidone (PVP, MW = 1,300,000, Aladdin), ascorbic acid (AA, Aldrich), and citriccid (CA, 99.9%, Aladdin) are used to obtain Ag NFs. Simply, a reaction vessel beaker is fixed in an ice-water bath environment. First, 10 mL deionized water was prepared in a 25 mL beaker with a magnetic stirrer. Then, $AgNO_3$ aqueous solution (1mL, 0.5 M), PVP solution (1 mL, 0.3 M), CA aqueous solution (0.1mL, 0.25 M) and AA aqueous solution (1 mL, 0.5 M) are separately added into deionized water (10 mL). Ag NFs are produced in a few seconds as our previous work[39]. Lastly, the fabricated Ag NFs are purified by acetone and ethyl alcohol through multiple centrifugation and dispersed in acetone for further usage.

*Random laser preparation*. First, the polydimethylsiloxane (PDMS, $n$=1.406) solution and the curing agent solution are mixed with a mass ratio of 10:1 to obtain a PDMS solution. Then 0.5 mL of dye-doped acetone solution (6 mg/mL DCJTB or 2 mg/mL PM567), 2 mL of PDMS solution and 0.2 mL of Ag NFs dispersion (5 mg/mL) are mixed and stirred, and then vacuumed to remove air bubbles in 30 minutes to obtain a mixture. Then, an appropriate amount of the mixture is dropped on a clean commercial optical fiber by using a micropipette. Due to gravity and surface tension, an asymmetric microcavity is formed, and microcavities with different sizes are made by controlling the intake of the pipette. At the same time, microcavities with the same gain dye or micro-cavities with different gain dyes can be made at different positions on the same optical fiber. Finally, the microcavities are placed in an 80 °C oven for 4 hours to get the random lasers.

**Random Lasing Characterization**. Scanning electron microscope (SEM, Hitachi SU8010) is used for material characterization. In our experiments, a Q-switched Nd: YAG laser is used as a pump source and supply 532 nm pulses with a pulse duration of 8 ns and the repetition frequency of 10 Hz. A cylindrical lens with a focal length $f$ = 10 cm is used to focus the laser beam and pump the microcavity for obtaining random lasing resonance. The AS distribution regulation is achieved through selectively pumping different microcavities by simply moving the mirror M2. The output random lasing can be divided into two channels for obtaining spectral information and AS distribution imaging information. An optical fiber spectrometer (Ocean Optics model Maya Pro 2000) with a resolution of 0.4 nm is used to detect the emission spectra. The used integration time is 100 ms. A camera (Nikon D5300, 6000*4000 pixels, 3.9 μm pixel size) is used for imaging. For demonstrating the potential of the designed random lasing in imaging and display, a broadband white source (Halogen Light Source, wave band:200~1700 nm) is used as an illumination source as comparisons.

**Data availability**

All data relevant to this work are available on request to the corresponding author.

**Conflicts of interest**

The authors declare no conflict of interest.

**Acknowledgements**

The authors thank the National Natural Science Foundation of China (grant no. 61975018, 11574033, and 11674032), the Beijing cooperative construction project, the Beijing Higher Education Young Elite Teacher Project and the Fundamental Research Funds for the Central Universities for financial support.

**Figure Captions:**

**Figure 1**

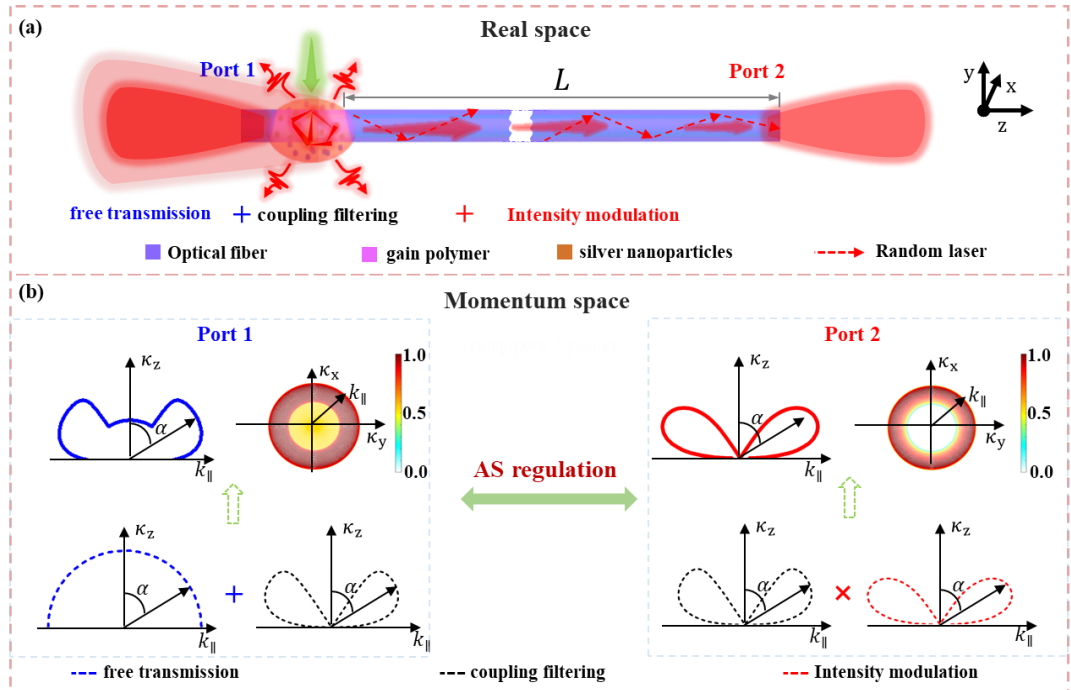

**Figure 1: Design mechanism of random lasers with tunable angular spectra. (a)** Random laser with adjustable angular spectra is designed with optical fiber in the real space; Port 1: emissions based on the free transmission and coupling filter; Port 2: emissions from coupling filtering and intensity modulation. **(b)** Angular spectra distribution of the random lasing in momentum space. Port 1: emissions including in-plane low spatial frequency angular spectrum (LSFAS); Port 2: emissions without in-plane LSFAS.

Figure 2

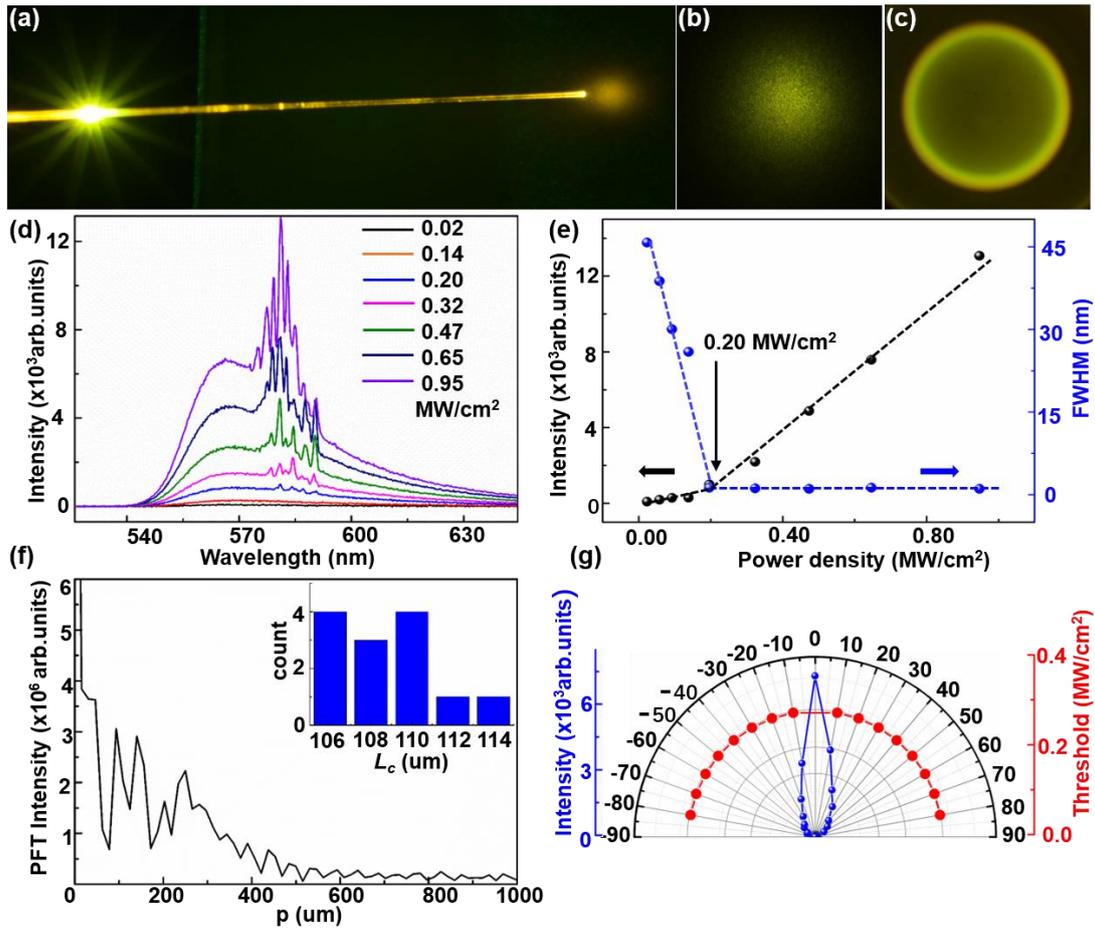

**Figure 2: Characterization of the output performance of the yellow random laser.** (**a-c**) optical photograph of random laser when PM567 asymmetric microcavity is pumped, photographs of random laser spots in real space (b) and momentum space (c). (**d**) The emission spectra of the random laser at various pump power densities. (**e**) The peak intensity and the half-height width of the laser mode centered at 580 nm vary with the pump power density, the threshold is 0.20 MW/cm$^2$. (**f**) The power Fourier transform of the yellow random laser spectra, the inset is the statistical distribution of the optical cavity length, indicating that the average effective cavity length is 108.68 μm. (**g**) For the microcavity with a 0.5 mg/mL concentration of PM567, the red line is the relationship between the measured threshold of the random laser and the pump angle. The blue line is the relationship between the peak intensity of the random laser and the detection angle, as pumping the asymmetric microcavity vertically.

**Figure 3**

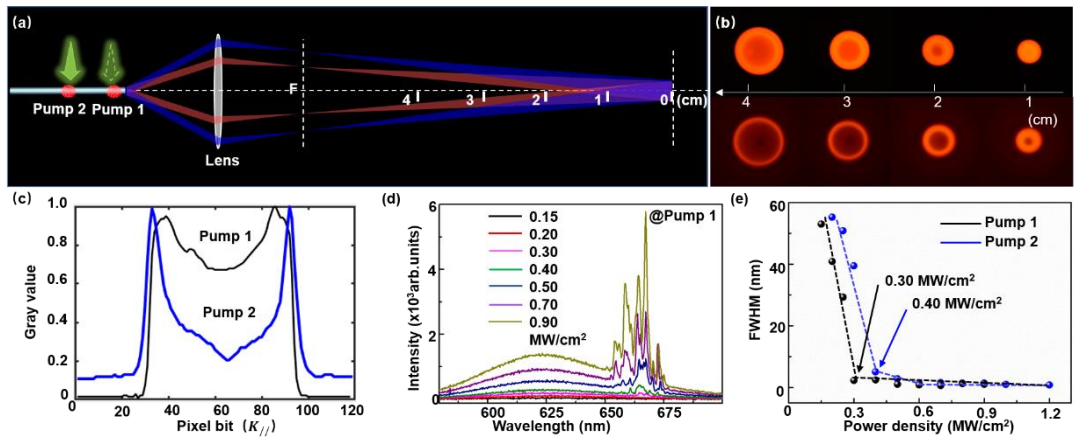

**Figure 3. Realization of the red random laser with tunable angular spectra (a)** The geometric optical transmission diagram of random lasing from the optical fiber port through the lens. **(b)** The intensity distribution diagram in the z-axis under pump mode 1 and pump mode 2, corresponding to the note in fig. a. The focal length of the selected lens is 3 cm, the fiber port is located 4cm in front of the lens, the origin is the position of the image. **(c)** the gray value distribution of the longitudinal diameter of the picture at -4 cm, with the pixel position changing. **(d)** The emission spectra of the random laser at various pump power densities, under 532 nm pulsed laser under pump mode 1. **(e)** The relationship between the half-height width of the laser mode centered at 660 nm with the pump power density, the thresholds under pump mode 1 and pump mode 2 are 0.30 MW/cm$^2$, 0.40 MW/cm$^2$.

Figure 4

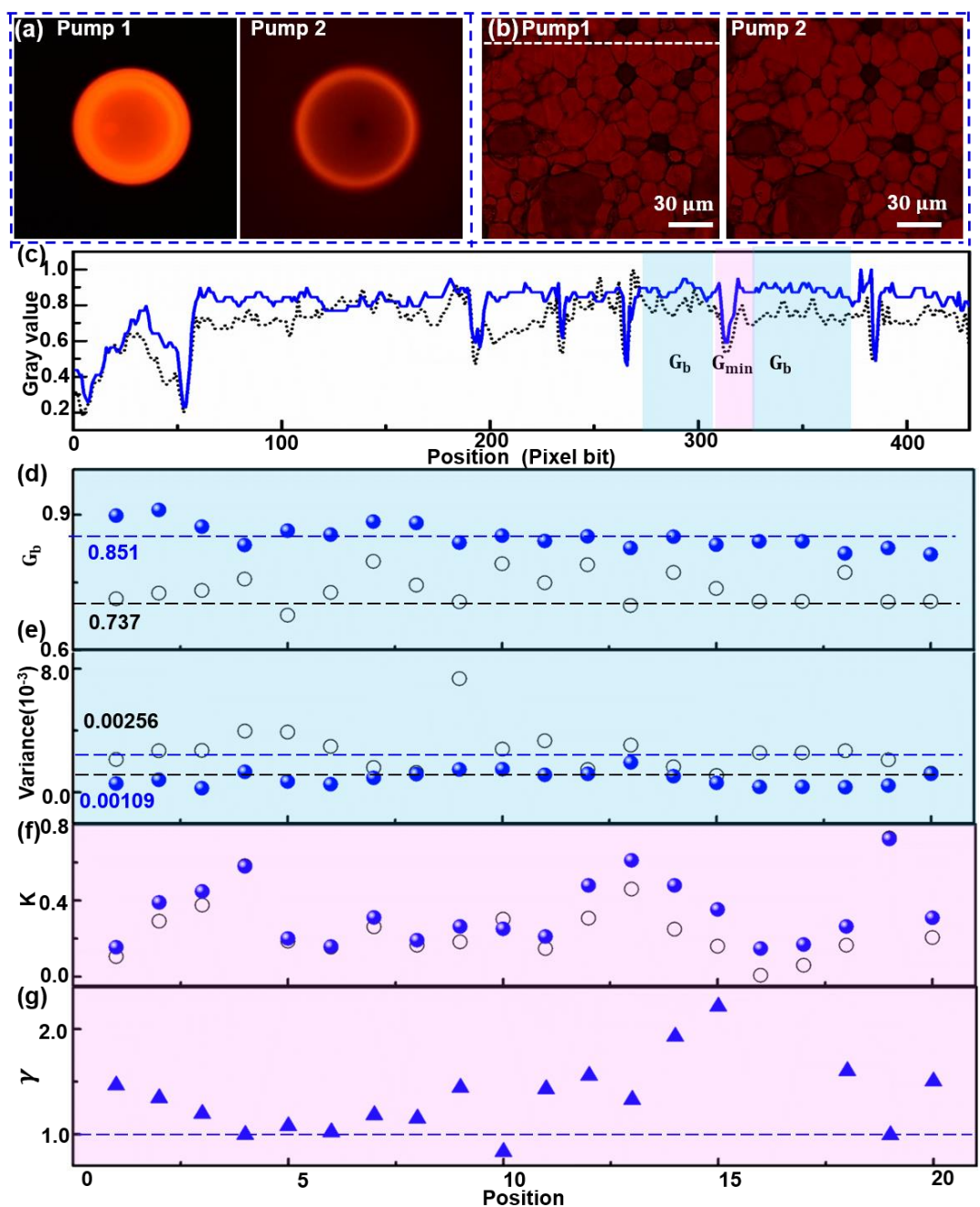

**Figure 4. Realization and application of random laser with adjustable angle spectrum. (a)** The spatial intensity distribution of the momentum measured by the camera in pump mode 1 and pump mode 2. **(b)** The microscopic imaging images of parenchyma cells under pump mode 1 and pump mode 2. **(c)** The gray value map of positions corresponding to the marks in the figure (b), which changes with the pixel position. The black dotted line and the blue solid line are the grayscale images of the random laser as the light source in pump mode 1 and pump mode 2. **(d)** The distribution

diagram of the background $G_b$, the variance, the modulation depth $K$ and improving factor $\gamma$ of the parenchyma cells. Corresponding to the $G_b$, take the corresponding the $G_{min}$ to obtain the distribution diagram of $K$ and $\gamma$. The white sphere is the data distribution graph of the random laser in pump mode 1, and the blue sphere is the data distribution graph of the random laser in pump mode 2. The blue triangle is the data distribution of the $k$ ratio ($K_{RL}$) / $K_{RL0}$) of the red random laser in pump mode 2 and mode 1.

**Figure 5**

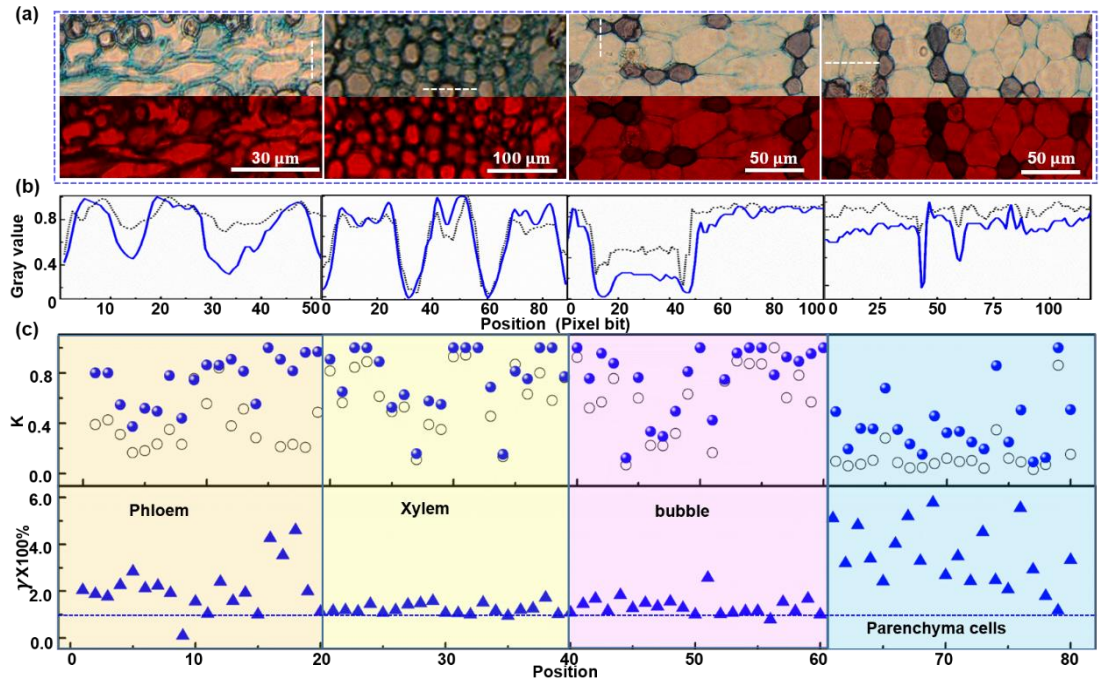

**Figure 5. Random laser modulation depth imaging application: white light and red random lasers are used as imaging light sources to partially image the cross-section slices of linden tree stem.** **(a)** Imaging picture of the partially enlarged locations of the phloem, xylem, bubble and parenchyma cells in the microscope imaging optical path, as white light and red random lasers are used as light sources. **(b)** The gray value map of positions corresponding to the marks in the graph (a), which changes with the pixel position. The Black dotted line and blue solid line are the gray maps of white light and red random laser as light sources. **(c)** Statistical data of the modulation depth K at the same position of the phloem, the xylem, the bubble and the

parenchyma cells. The white sphere is the data distribution of the white light K, and the blue sphere is the data distribution of the red random laser *K*. The blue triangle is data distribution of the k ratio ($K / K_0$) of the red random laser and the white light at different positions.